\documentclass{PoS}

\title{Lattice Simulations using OpenACC compilers}

\ShortTitle{OpenACC compilers}

\author{\speaker{Pushan Majumdar}\\
        Indian Association for the Cultivation of Science, Kolkata\\
        E-mail: \email{tppm@iacs.res.in}}

\abstract{OpenACC compilers allow one to use Graphics Processing Units without having to write 
explicit CUDA codes. Programs can be modified incrementally using OpenMP like directives which 
causes the compiler to generate CUDA kernels to be run on the GPUs. In this article we look at 
the performance gain in lattice simulations with dynamical fermions using OpenACC compilers.}

\FullConference{31st International Symposium on Lattice Field Theory - LATTICE 2013\\
		July 29 - August 3, 2013\\
		Mainz, Germany}

\begin{document}

\section{Introduction}
Graphics Processing Units (GPUs) are increasingly turning out to be the hardware of choice 
to run the compute intensive parts of lattice simulations since they have about an order 
of magnitude better price-performance ratio compared to normal CPUs. Additionally they 
consume only a fraction of the power for the same computing capacity.

Until recently, programs capable of running on GPUs had to be written in special 
languages (for example CUDA\cite{cuda}, which is like an extended version of C appropriate for 
the GPU) or even assembly level languages. A library of CUDA routines necessary for simulating 
lattice QCD is available \cite{quda}. However the need of a different language created a large 
barrier for a lot of people trying to use the GPUs.   

During the last two years there has been a development of directive based compilers 
called OpenACC compilers for both FORTRAN and C (much like OpenMP compilers) which 
allow programs written in standard FORTRAN and C to use the GPUs. In this article we 
report on our attempt to run lattice QCD simulation on GPUs using the Cray OpenACC 
FORTRAN compiler.

In section 2 we give a very brief introduction to OpenACC. Section 3 describes the 
matrix-vector multiplication routine which is called many times inside the conjugate 
gradient routine. Section 4 outlines the conjugate gradient subroutine on a single GPU 
while the next section points out the differences necessary for the multi-GPU version. Finally 
in section 6 we draw our conclusions. 

\section{OpenACC}
OpenACC is a programming standard for parallel computing developed by Cray, 
CAPS, Nvidia and PGI. The goal of the standard is to simplify parallel 
programming of heterogeneous CPU-accelerator systems \cite{wikipedia}.
The OpenACC Application Program Interface (API) lists a collection of compiler 
directives to specify loops and regions of code in standard C, C++ and Fortran 
programs which should run on an attached accelerator. The API is designed to 
provide portability across operating systems, host CPUs and accelerators.
The directives and programming model defined in this document allow programmers 
to create high-level host+accelerator programs without the need to explicitly initialize 
the accelerator, manage data or program transfers between the host and accelerator, 
or initiate accelerator startup and shutdown \cite{openacc}.

The different simulation programs we have had some experience with are :\\ 
(i) staggered fermions with the Wilson gauge action on both single and multiple GPUs,\\
(ii) Wilson fermions with the Wilson gauge action on a single GPU.

In the OpenACC programming model, even though the actual movement of the data is 
done automatically by the compiler, the programmer has to control the 
movement of the data carefully between the CPU and the GPU. This is crucial for efficient 
running of the program. The reason is the slow data movement between the CPU and the GPU. 
While the speed with which data is moved from the main memory of the computer to the 
CPU (with say 16 cores) is anything between 25 and 30 GB/s, the speed between the main 
memory and the GPU (with > 500 cores) which sits in a PCI slot is only about 5 GB/s. The 
result is if one is not careful with the movement of data, the GPU cores would spend most 
of their time waiting for the data.   

One option would be to bypass the CPU completely and do the computation entirely on the GPU. 
 However as of date it is impossible to do so as the GPU is incapable of 
handling input-output. Also in the openACC programming model, `if-then' clauses are evaluated on the CPU.
Moreover if one uses BLAS functions, they are launched from the CPU and MPI calls (at least for Fermi GPUs) are 
 also launched from the CPU.  

\section{Matrix-vector multiplication routine}
In this section we point out the differences between the openMP and openACC versions of the matrix-vector 
multiplication routine. One error we encountered with the presence of data while porting this routine 
to the GPU might be worth pointing out. The data is copied from the main memory to the GPU at the start 
of the conjugate gradient program. The matrix-vector multiplication routine uses the same data. Nevertheless 
we observed that if the compiler does not inline the routine, the matrix-vector multiplication routine does 
not find the necessary data. One simple way to avoid this problem is of course to inline the routine by 
hand, however that makes the program less modular and legible.

Below we list the differences between the openMP and openACC versions of the routine.\\  

\noindent
\verb|      subroutine matrix-vector(offset,local_length,v,w)| \\
\centerline{\ldots {\em All kinds of definitions and declarations} \ldots }

\noindent
{\em Original variable declaration in openMP program}\\
\verb|!$OMP parallel do default(shared)| \\
\verb|!$OMP+ private(nnu,px1,px2,px3,px4,px5,px6)| \\
\verb|!$OMP+ private(v1,v2,v3)|

\noindent {\em replaced by} 

\noindent
{\em modified variable declaration in openACC program}\\
\verb|!$ACC  parallel loop present(u,ud,v,w,iup,idn)| \\
\verb|!$ACC+ private(nnu,px1,px2,px3,px4,px5,px6,v1,v2,v3)|\\
\verb|!$ACC+ vector_length(32)|

\noindent
\verb|      do  l = offset+1, offset+local_length|\\
\hspace*{3cm}{\vdots} \\
\hspace*{2cm}{\em Routine identical to CPU version} \\
\hspace*{3cm}{\vdots} \\
\verb|      enddo| \\
\verb|      return|

The compiler generates the CUDA kernel for the loop immediately following the directives
 and launches it on the accelerator. The qualifier \verb|present(u,ud,v,w,iup,idn)| tells 
the compiler that the arrays \verb|u,ud,v,w,iup,idn| are already present on the accelerator.
Another possible option is to use \verb|present_or_copy|.
  
\section{Single GPU conjugate gradient routine}
In this section we outline the openACC version of the conjugate gradient 
routine which is the main work horse of the simulation program and takes between 80-90\% 
of the computing time. Therefore porting this routine to the GPU is essential for 
maximizing the gain from the GPU. 

The GPU portion of the code starts with a \verb|data| directive and ends with a 
\verb|end data| directive. Between these one can use other directives like 
\verb|copy,copyin,copyout,create| to 
move data between the CPU and the GPU or define variables on the GPU. As implied by the name, 
\verb|copyin| only copies the data into the GPU, \verb|copyout| copies the data from the GPU 
memory to the CPU memory, \verb|copy| copies the data from the CPU memory to the GPU memory 
at the beginning of the \verb|data| section and copies the updated value back to the CPU 
memory after the end of the \verb|data| section. \verb|create| defines the variables only 
on the GPU.

In the following code segment (in fortran) we often use $\langle \verb|x,y| \rangle$ to denote 
the inner product between \verb|x| and \verb|y| instead of writing out the loops.\\\\ 
\verb|       subroutine congrad(nitcg)| \\
\centerline{\ldots {\em All kinds of definitions and declarations} \ldots} \\
\vspace*{-5mm}
\begin{verbatim}
!$ACC  data copy(nitcg,alpha,betad,betan)
!$ACC+ copyin(nx,iup,idn,u,r)
!$ACC+ copyout(x,y)
!$ACC+ create(ud,ap,atap,p) 
      call linkc_acc
!$ACC  parallel loop collapse(2) reduction(+:betan) present(p,r,x)
      do l = 1, mvd2
      do ic=1,nc
         p(l,ic) = r(l,ic) ; x(l,ic) = (0.,0.)
         betan=betan+conjg(r(l,ic))*r(l,ic)
      end do
      end do
!$ACC  update host(betan)
      if (betan.lt.delit) go to 30
!$ACC  parallel present(beta,betan,betad,alphan)
       beta=betan/betad ; betad=betan ; alphan=betan
!$ACC  end parallel
\end{verbatim}
\verb|       do nx = 1, nitrc| \hfill{\em Main loop of conjugate gradient begins}\\
\verb|       nitcg = nitcg+1 ; ap = 0|  \\
\verb|       call fmv(0,mvd2,ap,p)| \hfill{\em fermion matrix on vector} \\
\verb|       alphad=|$\langle$\verb|ap,ap|$\rangle\, +\,\langle$\verb|p,p|$\rangle\quad;\quad $\verb|alpha=alphan/alphad| \\
\verb|       atap=p ; x = x + alpha*p| \\
\verb|       call fmtv(atap,ap)| \hfill{\em fermion matrix transpose on vector} \\
\verb|       r = r - alpha*atap| \\
\verb|       betan=|$\langle \verb|r,r| \rangle$ \\
\verb|!$ACC  update host(betan)       |\hfill {\em Exit condition evaluated on CPU} \\
\verb|       if (betan .lt. delit) go to 30| \\
\verb|       beta=betan/betad ; betad=betan ; alphan=betan| \\
\verb|       p = r + beta*p| \\
\verb|       end do       |\hfill{\em Main loop of conjugate gradient ends} \\
\verb|30     continue| \\
\verb|       y = 0        |\hfill{\em Solution on the second half lattice} \\
\verb|       call fmv(mvd2,mv,y,x)| \hfill{\em fermion matrix on vector}\\
\verb|!$ACC  end data| \\
\verb|       return|

The first new construct we encounter is the \verb|collapse(n)| directive. This directive directs the 
compiler to collapse the \verb|n| nested loops into a single parallel loop and create the kernel for 
that. For reduction operations, we need to specify the reduction variables to avoid error building.

The next new directive is \verb|upadte host|. As we have mentioned previously, the `if-then' branching 
conditions are evaluated on the CPU in the openACC programming model. Therefore if a variable is
 updated in the GPU memory and will be used in an `if-then' condition, then the updated variable 
has to be copied to the main memory. The \verb|update host| directive does just that. It updates the 
value of the variable in the argument of \verb|update host| in the main memory.

Even inside a data region, kernels for the GPU are created only when statements are within the 
\verb|parallel| and \verb|end parallel| directives. Otherwise they are evaluated on the CPU.
We find that the program runs faster if we give up the fortran vector and array constructs and perform
 operations on arrays using parallel loops. Our observation is that calling BLAS functions slows down the program 
a lot as they are launched only from the main memory. We therfore do all the copy and reduction operations 
using loops.

The compiler options (specific to the Cray openACC compiler) we used for the GPU portion are 
\verb| -hacc_model=fast_addr:auto_async_all|. 

\section{Multi-GPU conjugate gradient}
In this section we point out the changes that have to be made in the conjugate gradient and the matrix-vector 
multiplication routines (which in this case is manually inlined in the conjugate gradient) for the program to 
run on multiple GPUs which are on different nodes of a cluster and therefore have to use MPI calls to 
communicate between them. 

For simplicity, we assume in the following that only the fermion matrix on vector multiplication is split 
 between the nodes so that the output vector denoted by \verb|ap| in the previous section is replaced by 
\verb|ap_loc| and that needs to be communicated to the other nodes for constructing the full vector 
which will then serve as the input vector for the fermion matrix transpose on vector multiplication. \\\\
\verb|       subroutine congrad(nitcg)| \\
\centerline{\ldots {\em All kinds of definitions and declarations} \ldots} \\
\hspace*{3cm}{\vdots} \\
\hspace*{3cm} \verb|ap_loc = 0|
\begin{verbatim}
!$ACC  parallel loop present(u,ud,ap_loc,p,iup,idn)
      do  l = offset+1, offset+local_length
        v1 = ap_loc(1,l-offset)
\end{verbatim}
\hspace*{3cm}{\vdots} \\
\hspace*{2cm}{\em Lines identical to scalar version} \\
\hspace*{3cm}{\vdots}
\begin{verbatim}
        ap_loc(3,l-offset) = v3
      enddo
!$ACC update host(ap_loc)
      call MPI_ALLGATHER(ap_loc,3*local_length,MPI_DOUBLE_COMPLEX,
     +  ap,3*local_length,MPI_DOUBLE_COMPLEX,MPI_COMM_WORLD,ierr)
!$ACC  update device (ap)
\end{verbatim}
The new directive here is the \verb|update device|. After the local multiplication is 
done, the output is moved to the main memory by the \verb|update host(ap_loc)| directive.
 Then the \verb|MPI_ALLGATHER| is launched from the CPU to create the full input vector 
\verb|ap|. This updated \verb|ap| now has to be moved from the main memory to the GPU 
memory. This is achieved by the directive \verb|update device (ap)|.

The compiler option for the multi-GPU routine is slightly different. Instead of 
\verb|auto_async_all| we need to use \verb|auto_async_kernel| as otherwise the order 
of operations of \verb|update host|, \verb|MPI_ALLGATHER| and \verb|update device| 
will not be maintained.

\section{Performance \& Discussion}
\begin{figure}
\centerline{\includegraphics[width=0.58\textwidth,angle=-90]{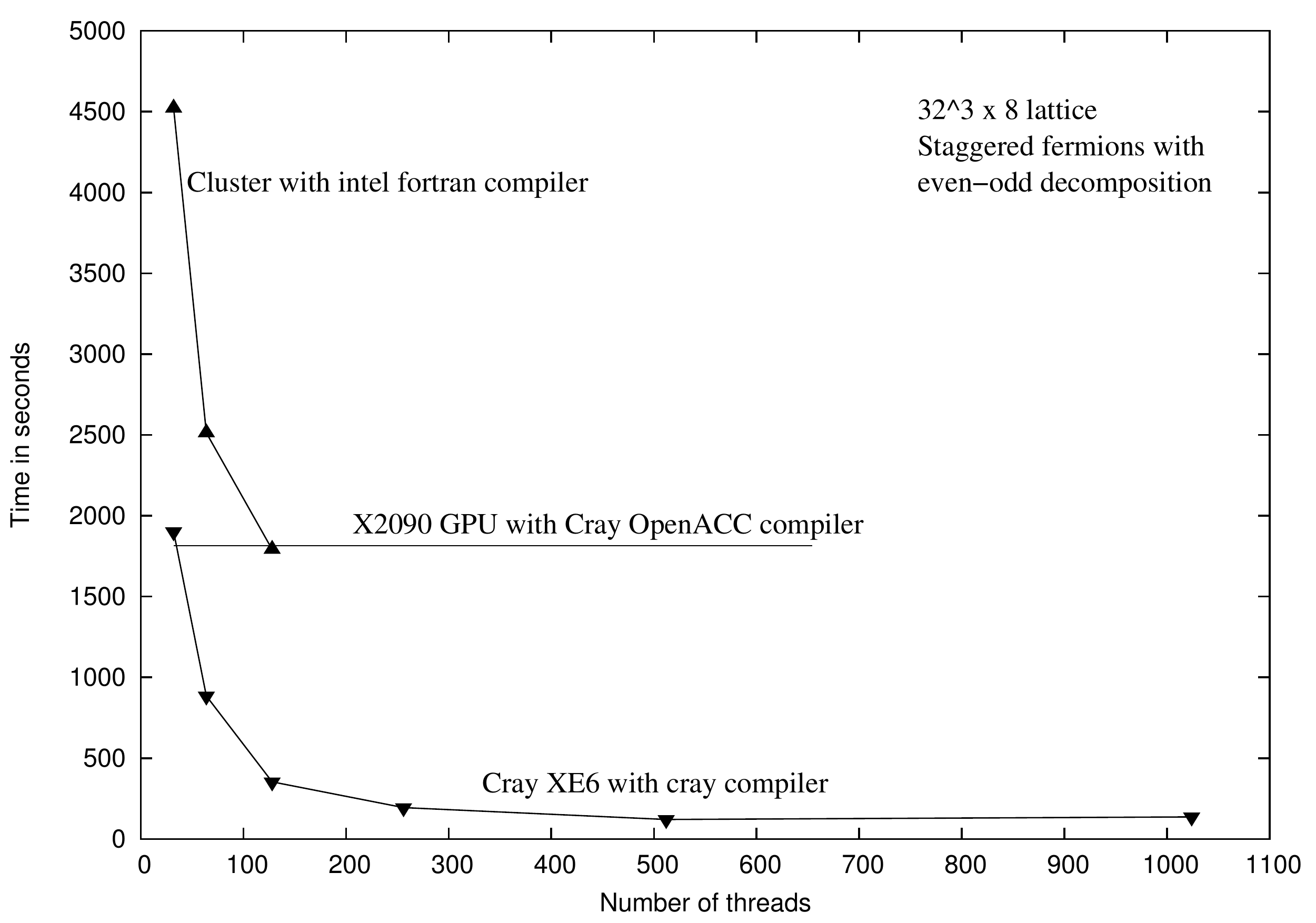}}
\caption{The figure compares the performance of a staggered fermion simulation program 
on a cluster (2.2 GHz Opteron (6 core) with QDR Infiniband interconnect) with the intel 
fortran compiler, a Cray XE6 (2.1 GHz Opteron (16 core) with Cray interconnect) with the 
cray fortran compiler and a single X2090 (Nvidia Fermi GPU) using the Cray OpenACC 
compiler.}  
\end{figure}

In figure 1 we compare the performance of a $N_f=2$ staggered fermion simulation program 
(Hybrid Molecular Dynamics) on a cluster, a Cray XE6 and X2090 GPU on one node of a Cray XK6. The 
 performance of single GPU staggered fermion code is roughly equivalent to 128 cores     
of cluster with QDR infiniband interconnect and slightly more than 32 cores of Cray 
XE6.

We also did some preliminary runs with a $N_f=2$ Wilson fermion program (Hybrid Monte Carlo). 
That gave a performance equivalent to about 96 cores of the cluster. For the Wilson fermion case, 
we have a hand coded CUDA program which is about 30\% faster than the openACC program.

Even without additional storage optimizations like storing only two 
columns of the gauge fields, a GPU with 6GB memory fits in a $32^4$ Wilson fermion lattice 
or a $10\times 40^3$ staggered fermion lattice.    

To summarize, in openACC programming, the coding effort is only marginally higher than 
OpenMP. Almost each OpenMP directive can be replaced with a OpenACC directive. The only 
additional structure is the creation of a data region with a list of variables (scalars 
and arrays) so that the compiler knows which variables to copy to the GPU and back again.
Since the connection between the GPU and main memory is comparatively slow, the data flow 
between the CPU and the GPU needs to be minimized. In fact we find an appreciable gain in 
performance only when the entire conjugate gradient routine runs on the GPU.

Fermi GPUs cannot issue MPI calls. So for Multi-GPU programs using MPI, every MPI call 
involves a data transfer from the GPU to CPU and back. Each such copy adds 
a significant ($\sim $ 15\%) overhead to the runtime. For Kepler GPUs the construct 
\verb|host_data use_device| cuts this down to a certain extent.
Nevertheless the timings we obtain show that GPUs are extremely powerful tools if one does 
not have access to conventional supercomputers.

\acknowledgments
The author would like to acknowledge
ILGTI-TIFR for funding the GPU portion of the Cray on which these studies were carried out and 
IACS for funding the rest of the machine. The author 
would also like to acknowledge the DST grant SR/S2/HEP-35/2008 for funding the cluster for the comparison runs 
and on which the hand coded CUDA program was developed. 
Finally the author also acknowledges the
Cray India team for help at various stages during the development of the OpenACC codes.

\end{document}